\documentclass[11pt]{article}
\usepackage{hyperref}
\pdfoutput=1

\begin{document}

\title{Unstable periodic orbits and heteroclinic\\connections in plane Couette flow}

\author{
J. F. Gibson$^1$, J. Halcrow$^1$, P. Cvitanovi\'c$^1$, and D. Viswanath$^2$\\
\\
$^1$School of Physics \\Georgia Institute of Technology \\Atlanta, GA  30332, USA\\
\\
$^2$Department of Mathematics\\ University of Michigan \\Ann Arbor, MI 48109, USA}

\maketitle

\begin{abstract}
Equilibrium, traveling wave, and periodic orbit solutions of pipe, channel, and plane Couette flows can now be computed precisely at Reynolds numbers above the onset of turbulence. These invariant solutions capture the complex dynamics of wall-bounded rolls and streaks and provide a framework for understanding low-Reynolds turbulent shear flows as dynamical systems. We present fluid dynamics videos of plane Couette flow illustrating periodic orbits, a close pass of turbulent flow to a periodic orbit, and heteroclinic connections between unstable equilibria.
\end{abstract}

\section{Description}

This article describes a \href{http://hdl.handle.net/1813/11491}{video
submitted to the 2008 Gallery of Fluid Motion}, for the annual meeting
of the American Physical Society's Division of Fluid Dynamics. The video
shows
\begin{itemize}
\item Examples of low-Reynolds turbulence in plane Couette flow, in periodic cells
with large and small aspect ratios.
\item Three periodic orbits of plane Couette flow in small periodic cells.
\item A close pass of a turbulent flow to a periodic orbit.
\item Two heteroclinic connections between unstable equilibria.
\end{itemize}

The animations feature a fluid visualization scheme explained
\href{http://cns.physics.gatech.edu/~gibson/PCF-movies/colorcoding.html}{here}.
All animations are direct numerical simulations of the Navier-Stokes equations
at $Re = 400$. Spatial discretization is Fourier $\times$ Chebyshev $\times$
Fourier, time-stepping is 3rd order semi-implicit backwards differentiation.
Geometrical parameters and further descriptive material is provided within the
video.

The equilibria are presented in \cite{HGC08},
the periodic orbits in \cite{GHC08a},
the heteroclinic connections in \cite{GHCV08},
and the state-space visualization method in \cite{GHC}.
The equilibria and periodic orbits were computed with a Newton-Krylov-hookstep
algorithm \cite{Viswanath1}.

Additional material can be found online at
\begin{itemize}
\item \href{http://www.channelflow.org}{www.channelflow.org}:
Software for direct numerical simulation and computation
of invariant solutions of plane COuette and channel flows.

\item \href{http://www.channelflow.org/database}{www.channelflow.org/database}:
Database of equilibria and periodic orbits of plane Couette flow.

\item \href{http://www.chaosbook.org}{www.chaosbook.org}:
Online dynamical systems textbook emphasizing unstable periodic orbits.

\item \href{http://www.chaosbook.org/tutorials}{www.chaosbook.org/tutorials}:
More animations.
\end{itemize}

\bibliographystyle{plain}
\bibliography{GibsonGFMentry}
\end{document}